\documentclass[12pt]{article}
\usepackage{amsmath}
\usepackage{amsfonts}
\usepackage{amssymb}
\usepackage[usenames]{color}
\usepackage{amsthm}

\def\Th{\Theta}

\def\p{\partial}
\def\e{\mathrm{e}}

\theoremstyle{remark}

\newcommand{\dbar}{\bar{\partial}}

\newcommand{\be}{\begin{equation}}
\newcommand{\ee}{\end{equation}}
\newcommand{\bea}{\begin{eqnarray}}
\newcommand{\eea}{\end{eqnarray}}
\newcommand{\beaa}{\begin{eqnarray*}}
\newcommand{\eeaa}{\end{eqnarray*}}

\newcommand{\nn}{\nonumber}
\renewcommand{\d}{\mathrm{d}}

\usepackage{authblk}
\begin{document}
\title
{Orlov-Schulman symmetries of the self-dual conformal structure equations
}
\author{L.V. Bogdanov\thanks{leonid@itp.ac.ru}}
\affil{Landau Institute for Theoretical Physics RAS}
\date{}
\maketitle
\begin{abstract}
We construct Orlov-Schulman symmetries for the self-dual conformal structure (SDCS) hierarchy.
We provide  an explicit proof of compatibility of additional symmetries with the basic Lax-Sato
flows of the  hierarchy, and  consider several simple examples, including 
Galilean transformations and scalings. 
We also present a picture of the Orlov-Schulman symmetries in terms of a dressing
scheme based on the Riemann-Hilbert problem. 
\end{abstract}
\section{Introduction}
This work can be seen as a continuation of the work \cite{LVB-OS25},
where we considered the (2+1)-dimensional case and constructed  Orlov-Schulman symmetries for  
the Manakov-Santini (MS) hierarchy.
In this work we  consider  four-dimentional case and construct Orlov-Schulman
symmetries for the self-dual conformal structure (SDCS) system hierarchy.

Orlov-Schulman symmetries were first introduced in the context of the
Ka\-dom\-tsev-Petviashvili (KP) hierarchy \cite{OS,Orlov88,GrinOrlov89}, they
form a Lie algebra of symmetries of the KP 
hierarchy, commuting with all the flows of the
hierarchy but not between themselves. Orlov-Schulman symmetries are currently widely used
in applications of the KP hierarchy to field theory, matrix integrals, string equations etc.
For the dispersionless limit of KP hierarchy (the dKP hierarchy) these symmetries were constructed
by Takasaki and Takebe \cite{Takasaki92}.

In the work \cite{LVB-OS25} we constructed Orlov-Schulman symmetries for the 
hierarchy of the Manakov-Santini (MS) system, 
which was introduced relatively recently as a generalisation of the dKP equation 
\cite{MS06, MS07, MS08}.
It was discovered \cite{DFK15} that this system possesses a deep geometric meaning,
describing a general local form of Einstein-Weyl spaces in (2+1) dimensions.

In this work we  consider  four-dimentional case and construct Orlov-Schulman
symmetries for the self-dual conformal structure (SDCS) system hierarchy.
Considering the Lax pair \cite{BDM06, BDM07}
\bea
\begin{aligned}
X_1&=\p_z-\lambda\p_x
+F_x\p_x+G_x\p_y+ f_x\p_\lambda,
\\
X_2&=\p_w- \lambda\p_y + F_y\p_x+G_y\p_y+f_y\p_\lambda.
\label{LaxSDCS}
\end{aligned}
\eea
we obtain a
coupled system of three second-order nonlinear PDEs for functions $F$, $G$, $f$
\bea
Q(F)=f_y, \quad Q(G)=-f_x,\quad Q(f)=0,
\label{SDCS3}
\eea
where a linear second order differential operator $Q$ is expressed as
\bea
\begin{aligned}
Q&=(\p_w+F_y\p_x+G_y\p_y)\p_x- (\p_z+F_x\p_x+G_x\p_y)\p_y
\\
&=\p_w\p_x-\p_z\p_y+F_y{\p_x}^2-G_x{\p_y}^2-
(F_x-G_y)\p_x\p_y.
\end{aligned}
\label{Q}
\eea
System (\ref{SDCS3}) can be rewritten in the form of a
coupled system of third order PDEs for the functions $F$, $G$
\bea
\left\{
\begin{aligned}
&
\p_x(Q(F))+\p_y(Q(G))=0,
\\
&
(\p_w+F_y\p_x+G_y\p_y)Q(G)+(\p_z+F_x\p_x+G_x\p_y)Q(F)=0,
\label{sd_3rd}
\end{aligned}
\right.
\label{SDCS2}
\eea
in this form it was introduced in \cite{DFK15}  as a general
local form of equations describing 
an (anti)self-dual conformal structure (SDCS) for the signature (2,\,2),
which is locally represented by the metric
\bea
\label{ASDmetric1}
g=dwdx-dzdy-F_y dw^2-(F_x-G_y)dwdz + G_xdz^2,
\eea
while the operator $Q$ defines a symmetric bivector corresponding to the conformal
structure. In recent work \cite{FK25} it was demonstrated that not only system (\ref{SDCS2}), but also the hierarchy for this system is connected  
with geometric structures, namely involutive scrolls. 
The work \cite{FK25} also considered another form of the SDCS equations, related to the Pleba\'nski first heavenly
equation, while system (\ref{SDCS2}) is related to the Pleba\'nski second  heavenly equation and 
contains it as a reduction. Indeed, after a volume-preserving reduction $F=\Theta_y$, $G=-\Theta_x$
we get 
Dunajski system generalising the second 
heavenly equation \cite{Dun02}
\bea
&&
\Th_{wx}-\Th_{zy}+\Th_{xx}\Th_{yy}-\Th_{xy}^2=f,
\nn\\
&&
\label{Dun}
f_{xw}-f_{yz}+
\Th_{yy}f_{xx}+\Th_{xx}f_{yy}-2\Th_{xy}f_{xy}=0,
\eea
and further reduction to linearly degenerate case
$f=0$ gives the famous Pleba\'nski second heavenly equation \cite{Pleb}
\bea
\Th_{wx}+\Th_{zy}+\Th_{xx}\Th_{yy}-\Th_{xy}^2=0.
\label{Heav}
\eea
The hierarchy and symmetries for this case were considered by Takasaki 
\cite{Takasaki89,Takasaki90}.

The SDCS hierarchy corresponds to the group of diffeomorphisms Diff(3)
with the symmetries represented by three-dimensiona vector fields,
The SDCS equations belongs to the class
of multidimensional dispersionless integrable systems having in generel no direct analogue in the
class of class of integrable systems with dispersion \cite{LVB09}. Thus the Orlov-Schulman
symmetries for this class cannot be obtained using the dispersionless limit.
However, the construction developed by Takasaki for the dKP hierarchy and Hamiltonian
vector fields (Poisson algebra) \cite{Takasaki92, Takasaki95}
can be extended to the case of general vector fields
and associated  multidimensional dispersionless integrable hierarchies.
Here we construct Orlov-Schulman symmetries for the SDCS hierarchy case, 
starting from the Lax-Sato equations of the hierarchy.
We also present a picture of Orlov-Schulman symmetries in terms of a dressing
scheme based on the Riemann-Hilbert problem \cite{LVB09}.
Similar to the KP hierarchy case, the simplest Orlov-Schulman symmetries correspond to scalings 
and Galilean transformations. For example, in terms of system (\ref{SDCS3})
we obtain the symmetries
\beaa
&&
f(\tau)= f_0( \e^\tau x, \e^\tau z; y, w),
\\
&&
F(\tau)=\e^{-\tau} F_0( \e^\tau x, \e^\tau z; y, w),
\\
&&
G(\tau)= G_0( \e^\tau x, \e^\tau z; y, w)
\eeaa
(asymmetric scaling) and
\beaa
&&
f(\tau)=f_0(x+\tau z, z; y+\tau w, w),
\\
&&
F(\tau)=F_0(x+\tau z, z; y+\tau w, w),
\\
&&
G(\tau)=G_0(x+\tau z, z; y+\tau w, w),
\eeaa
representing a symmetric Galilean transformation. There are also other types of Galilean transformations, which are introduced below.
\section{The SDCS system hierarchy}
Lax-Sato equations of the SDCS hierarchy (see \cite{BDM07,LVB09,LVB25})
\bea
\partial^1_n\mathbf{\Psi}=
+\left(
\frac{(\Psi^0)^{n}}{J_0}
\begin{vmatrix}
\Psi^0_\lambda & \Psi^2_\lambda\\
\Psi^0_y & \Psi^2_y
\end{vmatrix}
\right)_+\partial_x \mathbf{\Psi}-
\left(\frac{(\Psi^0)^{n}}{J_0}
\begin{vmatrix}
\Psi^0_\lambda & \Psi^2_\lambda\\
\Psi^0_x & \Psi^2_x
\end{vmatrix}
\right)_+\partial_y \mathbf{\Psi}-
\nn\\
\left(
\frac{(\Psi^0)^{n}}{J_0}
\begin{vmatrix}
\Psi^0_x & \Psi^2_x\\
\Psi^0_y & \Psi^2_y
\end{vmatrix}
\right)_+\partial_\lambda \mathbf{\Psi}
,
\label{SD1}
\eea
\bea
\partial^2_n\mathbf{\Psi}=
-\left(\frac{(\Psi^0)^{n}}{J_0}
\begin{vmatrix}
\Psi^0_\lambda & \Psi^1_\lambda\\
\Psi^0_y & \Psi^1_y
\end{vmatrix}
\right)_+\partial_x \mathbf{\Psi}+
\left(\frac{(\Psi^0)^{n}}{J_0}
\begin{vmatrix}
\Psi^0_\lambda & \Psi^1_\lambda\\
\Psi^0_x & \Psi^1_x
\end{vmatrix}
\right)_+\partial_y \mathbf{\Psi}+
\nn\\
\left(\frac{(\Psi^0)^{n}}{J_0}
\begin{vmatrix}
\Psi^0_x & \Psi^1_x\\
\Psi^0_y & \Psi^1_y
\end{vmatrix}
\right)_+\partial_\lambda \mathbf{\Psi},
\label{SD2}
\eea
where $\mathbf{\Psi}=(\Psi^0, \Psi^1, \Psi^2)$, 
$\partial^k_n=\frac{\partial}{\partial t^k_n}$, $|A|=\det(A)$,
define the evolution of three formal series 
\bea
&&
\Psi^0=\lambda+\sum_{n=1}^\infty \Psi^0_n(\mathbf{t}^1,\mathbf{t}^2)\lambda^{-n},
\label{form0}
\\&&
\Psi^1=\sum_{n=0}^\infty t^1_n (\Psi^0)^{n}+
\sum_{n=1}^\infty \Psi^1_n(\mathbf{t}^1,\mathbf{t}^2)\lambda^{-n}
\label{form1}
\\&&
\Psi^2=\sum_{n=0}^\infty t^2_n (\Psi^0)^{n}+
\sum_{n=1}^\infty \Psi^2_n(\mathbf{t}^1,\mathbf{t}^2)\lambda^{-n}
\label{form2}
\eea 
with respect to higher times of the hierarchy.
Here $\mathbf{t}^1=(t^1_0,\dots,t^1_n,\dots)$, $\mathbf{t}^2=(t^2_0,\dots,t^2_n,\dots)$,
$x=t^1_0$, $y=t^2_0$. 
We use the projectors $(\sum_{-\infty}^{\infty}u_n \lambda^n)_+
=\sum_{n=0}^{\infty}u_n \lambda^n$,
$(\sum_{-\infty}^{\infty}u_n \lambda^n)_-=\sum_{-\infty}^{n=-1}u_n \lambda^n$,
$J_0$ is the determinant of the Jacobian matrix,
\be
J_0=
\begin{vmatrix}
\Psi^0_\lambda &\Psi^1_\lambda&\Psi^2_\lambda \\
\Psi^0_x&\Psi^1_x &\Psi^2_x \\
\Psi^0_y&\Psi^1_y &\Psi^2_y
\end{vmatrix}.
\label{J0}
\ee
The first two flows of the hierarchy (\ref{SD1}), (\ref{SD2})
read
\beaa
&&\p_1^1\mathbf{\Psi}=(\lambda\p_x
-F_{x}\p_x-G_{x}\p_y-f_{x}\p_{\lambda})\mathbf{\Psi},
\\
&&\p_1^2\mathbf{\Psi}=(\lambda\p_y
-F_{y}\p_x
-G_{y}\p_y -f_{y}\p_{\lambda})\mathbf{\Psi},
\eeaa
where
$$
G=\Psi^2_1,\quad F=\Psi^1_1, \quad f=\Psi^0_1,
$$
and after the identification $z=t^1_1$, $w=t^2_1$ we obtain linear equations 
for Lax pair (\ref{LaxSDCS}) with  $\Psi^0, \Psi^1, \Psi^2$  
playing a role of the basic wave functions.

In short form, we write equations (\ref{SD1}), (\ref{SD2}) as follows:
\bea
\partial^1_n\mathbf{\Psi}={\hat V}^{1}_{n+}\mathbf{\Psi},
\quad
\partial^2_n\mathbf{\Psi}={\hat V}^{2}_{n+}\mathbf{\Psi},
\label{SDc}
\eea
where the vector fields ${\hat V}^{1}_{n+}$,  ${\hat V}^{2}_{n+}$ are defined by 
the r.h.s. of equations (\ref{SD1}), (\ref{SD2}). Below we also use the vector fields
${\hat V}^{1}_{n-}$,  ${\hat V}^{2}_{n-}$  and  ${\hat V}^{1}_{n}$,  ${\hat V}^{2}_{n}$, where
the coefficients are used with minus projection or without projection. 
Evidently,
${\hat V}^{1}_{n}={\hat V}^{1}_{n+}+{\hat V}^{1}_{n-}$, 
${\hat V}^{2}_{n}={\hat V}^{2}_{n+}+{\hat V}^{2}_{n-}$. 
We also have a simple identity that follows directly from the definition of vector fields,
\beaa
{\hat V}^{\alpha}_{n}\Psi^i=\delta^\alpha_i(\Psi^0)^{n},\quad \alpha=1;2,\quad i=0;1;2,
\quad 1\leqslant n\leqslant \infty.
\eeaa

\section{Orlov-Schulman symmetries}
We introduce Orlov-Schulman symmetries of the SDCS hierarchy through some special `minus'
vector fields (compare \cite{Takasaki92}, \cite{LVB-OS25})
\bea
\partial_\tau\mathbf{\Psi}=
-\left(
\frac{\Phi^0}{J_0}
\begin{vmatrix}
\Phi^1_\lambda & \Phi^2_\lambda\\
\Phi^1_y & \Phi^2_y
\end{vmatrix}
\right)_-
\partial_x \mathbf{\Psi}
+\left(\frac{\Phi^0}{J_0}
\begin{vmatrix}
\Phi^1_\lambda & \Phi^2_\lambda\\
\Phi^1_x & \Phi^2_x
\end{vmatrix}
\right)_-
\partial_y \mathbf{\Psi}
\nn\\
+\left(
\frac{\Phi^0}{J_0}
\begin{vmatrix}
\Phi^1_x & \Phi^2_x\\
\Phi^1_y & \Phi^2_y
\end{vmatrix}
\right)_-
\partial_\lambda \mathbf{\Psi}
,
\label{OSsd}
\eea
where $\Phi^0$, $\Phi^1$, $\Phi^2$ are some solutions of linear equations of the SDCS hierarchy,
\beaa
{\hat V}^{\alpha}_{n+}\Phi^i=0,\quad \alpha=1;2,\quad i=0;1;2, \quad 1\leqslant n\leqslant \infty.
\eeaa
A general solution of these equations is expressed as an arbitrary holomorphic function 
of three basic functions $\Phi=F(\Psi^0,\Psi^1,\Psi^2)$.
In a concise form, we will write symmetry (\ref{OSsd}) as
\beaa
\partial_\tau \mathbf{\Psi}=-{\hat U}_{-}\mathbf{\Psi}.
\eeaa
For $\Phi^0=1$ the flow is compatible with the volume-preserving reduction of the hierarchy 
$J_0=1$.

Using an identity
\beaa
-\left(
\frac{\Phi^0}{J_0}
\begin{vmatrix}
\Phi^1_\lambda & \Phi^2_\lambda\\
\Phi^1_y & \Phi^2_y
\end{vmatrix}
\right)
\partial_x {\Psi^i}
+\left(\frac{\Phi^0}{J_0}
\begin{vmatrix}
\Phi^1_\lambda & \Phi^2_\lambda\\
\Phi^1_x & \Phi^2_x
\end{vmatrix}
\right)
\partial_y {\Psi^i}
+\left(
\frac{\Phi^0}{J_0}
\begin{vmatrix}
\Phi^1_x & \Phi^2_x\\
\Phi^1_y & \Phi^2_y
\end{vmatrix}
\right)
\partial_\lambda {\Psi^i}
\nn\\
=
\frac{\Phi^0}{J_0}\left|\frac{\p(\Phi^1,\Phi^2,\Psi^i)}{\p(\lambda,x,y)}\right|
\eeaa
or, in short,
\beaa
-{\hat U} {\Psi}^i=\psi^i= 
\frac{\Phi^0}{J_0}\left|\frac{\p(\Phi^1,\Phi^2,\Psi^i)}{\p(\lambda,x,y)}\right|
\eeaa
in terms of `plus'  vector fields we obtain
\bea
\partial_\tau{\Psi^i}=
+\left(
\frac{\Phi^0}{J_0}
\begin{vmatrix}
\Phi^1_\lambda & \Phi^2_\lambda\\
\Phi^1_y & \Phi^2_y
\end{vmatrix}
\right)_+
\partial_x {\Psi^i}
-\left(\frac{\Phi^0}{J_0}
\begin{vmatrix}
\Phi^1_\lambda & \Phi^2_\lambda\\
\Phi^1_x & \Phi^2_x
\end{vmatrix}
\right)_+
\partial_y {\Psi^i}
\nn\\
-\left(
\frac{\Phi^0}{J_0}
\begin{vmatrix}
\Phi^1_x & \Phi^2_x\\
\Phi^1_y & \Phi^2_y
\end{vmatrix}
\right)_+
\partial_\lambda {\Psi^i}
+
\frac{\Phi^0}{J_0}\left|\frac{\p(\Phi^1,\Phi^2,\Psi^i)}{\p(\lambda,x,y)}\right|,
\label{OSsd+}
\eea
the last term can be rewritten in the form
\beaa
\psi^i=\frac{\Phi^0}{J_0}\left|\frac{\p(\Phi^1,\Phi^2,\Psi^i)}{\p(\lambda,x,y)}\right|
=\Phi^0 \left|\frac{\p(\Phi^1,\Phi^2,\Psi^i)}{\p(\Psi^0,\Psi^1,\Psi^2)}\right|,
\eeaa
in which it is evident that it is a wave function 
of linear equations of the hierarchy (\ref{SDc}), and the short form of (\ref{OSsd+})
is
\bea
\partial_\tau{\Psi^i}=\hat U_+ {\Psi^i} + \psi^i.
\label{OSc+}
\eea
\section{Compatibility of Orlov-Schulman symmetries with the hierarchy}
A special structure of the Orlov-Schulman symmetries (\ref{OSsd}) 
guarantees their compatibility with Lax-Sato flows  (\ref{SD1}), (\ref{SD2})
of the SDCS hierarchy. For the MS hierarchy case the proof of compatibility 
was given in the work \cite{LVB-OS25}, here it is very similar.
The flows of the hierarchy and an extra symmetry
in terms of `plus' vector fields read
\beaa
&&
\partial^\alpha_n{\Psi^i}={\hat V}^{\alpha}_{n+}{\Psi^i},
\\
&&
\partial_\tau{\Psi^i}=\hat U_+ {\Psi^i} + \psi^i
\eeaa
At this stage we consider  $\partial_\tau$ as infinitesimal symmetry, which in general
may not commute with the flows of the hierarchy,
and define a set of commutators
\beaa
\Delta^\alpha_n(i)=\partial^\alpha_n \partial_\tau{\Psi^i}
-\partial_\tau \partial^\alpha_n{\Psi^i},
\eeaa
which are equal to
\beaa
\Delta^\alpha_n(i)=\partial^\alpha_n 
( \hat U_+ {\Psi^i} + \psi^i)-\partial_\tau ({\hat V}^{\alpha}_{n+}{\Psi^i})
\eeaa
or
\beaa
\Delta^\alpha_n(i)=(\partial^\alpha_n\hat U_+ +\hat U_+{\hat V}^{\alpha}_{n+}){\Psi^i}
+ \partial^\alpha_n  \psi^i - (\partial_\tau {\hat V}^{\alpha}_{n+} + 
{\hat V}^{\alpha}_{n+}\hat U_+){\Psi^i} - {\hat V}^{\alpha}_{n+}\psi^i
\eeaa
Taking into account that functions $\psi^i$ are solutions of linear equations of the SDCS hierarchy,
we get
\beaa
\Delta^\alpha_n(i)=\left(\partial^\alpha_n\hat U_+ 
-  \partial_\tau {\hat V}^{\alpha}_{n+}   
+     [\hat U_+;{\hat V}^{\alpha}_{n+} ]
\right){\Psi^i}
\eeaa

Using equivalent representation of the flows of the hierarchy and extra symmetry
in terms of `minus' vector fields
\beaa
&&
\partial^\alpha_n{\Psi^i}=-{\hat V}^{\alpha}_{n-}{\Psi^i} + \delta^\alpha_i(\Psi^0)^{n},
\\
&&
\partial_\tau{\Psi^i}=-\hat U_- {\Psi^i}, 
\eeaa
for the commutators we obtain
\beaa
\Delta^\alpha_n(i)=-\partial^\alpha_n 
( \hat U_- {\Psi^i} )+\partial_\tau ({\hat V}^{\alpha}_{n-}{\Psi^i}- \delta^\alpha_i(\Psi^0)^{n}),
\eeaa
and, after some transformations,
\beaa
\Delta^\alpha_n(i)=-\left(\partial^\alpha_n\hat U_- 
-  \partial_\tau {\hat V}^{\alpha}_{n-}   
+     [\hat U_-;{\hat V}^{\alpha}_{n-} ]
\right){\Psi^i}
\eeaa
Taking into account similar expression in terms of the `plus' vector fields,
we obtain an identity
\beaa
\left((\partial^\alpha_n\hat U_+ 
-  \partial_\tau {\hat V}^{\alpha}_{n+}   
+     [\hat U_+;{\hat V}^{\alpha}_{n+} ]) + 
(\partial^\alpha_n\hat U_- 
-  \partial_\tau {\hat V}^{\alpha}_{n-}   
+     [\hat U_-;{\hat V}^{\alpha}_{n-} ])
\right){\Psi^i}=0, 
\eeaa
where $i=0;1;2$,
meaning that a vector field (containing derivatives $\p_\lambda$, $\p_x$, $\p_y$) annulates a set of
three basic functions $\Psi^0$,  $\Psi^1$,  $\Psi^2$ (with nonzero Jacobian $J_0$); then
this vector field is identically zero. Taking `plus' and `minus' projections of this vector field, we come
to the conclusion that $\Delta^\alpha_n(i)=0$ and an extra symmetry commutes with all the flows of
the hierarchy, Q.E.D.
\section{Examples} 
\subsection{Scaling transformations}
We start from the representation of Orlov-Schulman symmetries 
in terms of `plus' vector fields (\ref{OSsd+}), (\ref{OSc+}).
\paragraph*{Example 1}
Let us take $\Phi^1=\Psi^0$, $\Phi^2=\Psi^2$, $\Phi^0=\Psi^1$,
then 
\beaa
\psi^0=0, \quad \psi^1=-\Psi^1,\quad \psi^2=0.
\eeaa
Comparing equation (\ref{OSsd+}) with Lax-Sato equations (\ref{SD1}), (\ref{SD2}),
we obtain
\beaa
\partial_\tau 
\begin{pmatrix}
\Psi^0
\\
\Psi^1
\\
\Psi^2
\end{pmatrix}=\sum t^1_n \p^1_n  
\begin{pmatrix}
\Psi^0
\\
\Psi^1
\\
\Psi^2
\end{pmatrix}
+ 
\begin{pmatrix}
0
\\
-\Psi^1
\\
0
\end{pmatrix},
\eeaa
and in terms of the functions $f$, $F$, $G$ entering SDCS system (\ref{SDCS3})
we get the symmetry 
\beaa
\partial_\tau f=\sum t^1_n \p^1_n f,
\quad
\partial_\tau F=\sum t^1_n \p^1_n  F - F,
\quad
\partial_\tau G=\sum t^1_n \p^1_n  G,
\eeaa
which can be explicitely integrated and gives a scaling transformation
for the first half of independent variables,
\beaa
f(\tau)= f_0( \e^\tau\mathbf{t}^1;\mathbf{t}^2),
\quad
F(\tau)=\e^{-\tau} F_0( \e^\tau\mathbf{t}^1;\mathbf{t}^2),
\quad
G(\tau)= G_0( \e^\tau\mathbf{t}^1;\mathbf{t}^2).
\eeaa
For the Dunajski system (\ref{Dun})
the scaling takes the form
\beaa
&&
\Theta(\tau)=\e^{-\tau} \Theta_0( \e^\tau x, \e^\tau z,  ;y, w),
\\
&&
f(\tau)=f_0( \e^\tau x, \e^\tau z,  ;y, w),
\eeaa
respectively for the first heavenly equation $f=0$ and we have a scaling for $\Theta$.
\paragraph*{Example 2}
Similarly, for $\Phi^1=\Psi^0$, $\Phi^2=\Psi^1$, $\Phi^0=\Psi^2$
we have
\beaa
\partial_\tau 
\begin{pmatrix}
\Psi^0
\\
\Psi^1
\\
\Psi^2
\end{pmatrix}=-\sum t^2_n \p^2_n  
\begin{pmatrix}
\Psi^0
\\
\Psi^1
\\
\Psi^2
\end{pmatrix}
+ 
\begin{pmatrix}
0
\\
0
\\
\Psi^2
\end{pmatrix},
\eeaa
leading to the scaling of the second half of independent variables
\beaa
f(\tau)= f_0(\mathbf{t}^1;\e^{-\tau} \mathbf{t}^2),
\qquad
F(\tau)=F_0( \mathbf{t}^1;\e^{-\tau} \mathbf{t}^2),
\qquad
G(\tau)=\e^{\tau}  G_0( \mathbf{t}^1;\e^{-\tau} \mathbf{t}^2).
\eeaa
\paragraph*{Example 3}Linear combination of two preceding cases
$
\p_\tau=\p_{\tau_1}+ \p_{\tau_2}
$
corresponds to antisymmetric scaling
\beaa
f(\tau)= f_0(\e^{\tau}\mathbf{t}^1;\e^{-\tau} \mathbf{t}^2),
\quad
F(\tau)=\e^{-\tau} F_0(\e^{\tau} \mathbf{t}^1;\e^{-\tau} \mathbf{t}^2),
\quad
G(\tau)=\e^{\tau} G_0( \e^\tau\mathbf{t}^1;\e^{-\tau} \mathbf{t}^2).
\eeaa
\paragraph*{Example 4}
For
$
\p_\tau=\p_{\tau_1}- \p_{\tau_2}
$
we obtain homogeneous scaling
\beaa
f(\tau)= f_0(\e^{\tau}\mathbf{t}^1;\e^{\tau} \mathbf{t}^2),
\quad
F(\tau)=\e^{-\tau} F_0(\e^{\tau} \mathbf{t}^1;\e^{\tau} \mathbf{t}^2),
\quad
G(\tau)=\e^{-\tau} G_0( \e^\tau\mathbf{t}^1;\e^{\tau} \mathbf{t}^2).
\eeaa
\subsection{Galilean transformations and rotations}
\paragraph*{Example 1} Let $\Phi^1=\Psi^0$, $\Phi^2=\Psi^2$, $\Phi^0=\Psi^2$, then
\beaa
\partial_\tau 
\begin{pmatrix}
\Psi^0
\\
\Psi^1
\\
\Psi^2
\end{pmatrix}=\sum t^2_n \p^1_n  
\begin{pmatrix}
\Psi^0
\\
\Psi^1
\\
\Psi^2
\end{pmatrix}
+ 
\begin{pmatrix}
0
\\
-\Psi^2
\\
0
\end{pmatrix},
\eeaa
resulting in an asymmetric Galilean transformation
\beaa
&&
f(\tau)= f( \mathbf{t}^1+\tau \mathbf{t}^2 ;\mathbf{t}^2),
\\
&&
F(\tau)=F_0( \mathbf{t}^1+\tau \mathbf{t}^2 ;\mathbf{t}^2)
- \tau G_0( \mathbf{t}^1+\tau \mathbf{t}^2 ;\mathbf{t}^2) ,
\\
&&
G(\tau)= G_0( \mathbf{t}^1+\tau \mathbf{t}^2 ;\mathbf{t}^2).
\eeaa
For the Dunajski system (\ref{Dun})
\beaa
\Theta(\tau)=\Theta_0(x+\tau y, z+\tau w; y, w),
\quad
f(\tau)=f_0(x+\tau y, z+\tau w; y, w).
\eeaa
\paragraph*{Example 2} In complete analogy, for $\Phi^1=\Psi^0$, $\Phi^2=\Psi^1$, $\Phi^0=\Psi^1$ we have
\beaa
\partial_\tau 
\begin{pmatrix}
\Psi^0
\\
\Psi^1
\\
\Psi^2
\end{pmatrix}=-\sum t^1_n \p^2_n  
\begin{pmatrix}
\Psi^0
\\
\Psi^1
\\
\Psi^2
\end{pmatrix}
+ 
\begin{pmatrix}
0
\\
0
\\
\Psi^1
\end{pmatrix}
\eeaa
resulting in antisymmetric Galilean transformation
\beaa
&&
f(\tau)= f_0(\mathbf{t}^1;  \mathbf{t}^2-\tau \mathbf{t}^1 ),
\\
&&
F(\tau)=F_0(\mathbf{t}^1;  \mathbf{t}^2-\tau \mathbf{t}^1 )
,
\\
&&
G(\tau)= G_0(\mathbf{t}^1;  \mathbf{t}^2-\tau \mathbf{t}^1 )
+\tau F_0(\mathbf{t}^1;  \mathbf{t}^2-\tau \mathbf{t}^1 ).
\eeaa
For Dunajski system (\ref{Dun})
\beaa
\Theta(\tau)=\Theta_0(x, z; y-\tau x, w-\tau z),
\quad
f(\tau)=f_0(x, z; y-\tau x, w-\tau z).
\eeaa
\paragraph*{Example 3} Linear combinations of flows of  Ex.1 and Ex.2.
The sum of generators
$
\p_\tau=\p_{\tau_1}+ \p_{\tau_2}
$
corresponds to rotation
\beaa
&&
f(\tau)= f_0(\cos(\tau)\mathbf{t}^1 +\sin(\tau)\mathbf{t}^2;  
\cos(\tau)\mathbf{t}^2 - \sin(\tau)\mathbf{t}^1),
\\
&&
F(\tau)=
\cos(\tau)F_0(\cos(\tau)\mathbf{t}^1 
+\sin(\tau)\mathbf{t}^2;  
\cos(\tau)\mathbf{t}^2 - \sin(\tau)\mathbf{t}^1)
\\
&&\qquad\qquad
-
\sin(\tau)G_0(\cos(\tau)\mathbf{t}^1 +\sin(\tau)\mathbf{t}^2;  
\cos(\tau)\mathbf{t}^2 - \sin(\tau)\mathbf{t}^1)
,
\\
&&
G(\tau)=
\cos(\tau)G_0(\cos(\tau)\mathbf{t}^1 +\sin(\tau)\mathbf{t}^2;  
\cos(\tau)\mathbf{t}^2 - \sin(\tau)\mathbf{t}^1)
\\
&&\qquad\qquad
+
\sin(\tau)F_0(\cos(\tau)\mathbf{t}^1 +\sin(\tau)\mathbf{t}^2;  
\cos(\tau)\mathbf{t}^2 - \sin(\tau)\mathbf{t}^1).
\eeaa
For the potential $\Theta$
\beaa
\Theta(\tau)=\Theta_0(\cos(\tau)\mathbf{t}^1 +\sin(\tau)\mathbf{t}^2;  
\cos(\tau)\mathbf{t}^2 - \sin(\tau)\mathbf{t}^1)
\eeaa
\paragraph*{Example 4}
The difference of generators
$
\p_\tau=\p_{\tau_1} - \p_{\tau_2}
$
corresponds to hyperbolic rotation
\beaa
&&
f(\tau)= f_0(\cosh(\tau)\mathbf{t}^1 + \sinh(\tau)\mathbf{t}^2;  
\cosh(\tau)\mathbf{t}^2 + \sinh(\tau)\mathbf{t}^1),
\\
&&
F(\tau)=
\cosh(\tau)F_0(\cosh(\tau)\mathbf{t}^1 
+\sin(\tau)\mathbf{t}^2;  
\cos(\tau)\mathbf{t}^2 + \sin(\tau)\mathbf{t}^1)
\\
&&\qquad\qquad
-
\sinh(\tau)G_0(\cosh(\tau)\mathbf{t}^1 +\sinh(\tau)\mathbf{t}^2;  
\cosh(\tau)\mathbf{t}^2 +\sinh(\tau)\mathbf{t}^1)
,
\\
&&
G(\tau)=
\cosh(\tau)G_0(\cosh(\tau)\mathbf{t}^1 +\sinh(\tau)\mathbf{t}^2;  
\cosh(\tau)\mathbf{t}^2 + \sinh(\tau)\mathbf{t}^1)
\\
&&\qquad\qquad
-
\sinh(\tau)F_0(\cosh(\tau)\mathbf{t}^1 +\sinh(\tau)\mathbf{t}^2;  
\cos(\tau)\mathbf{t}^2 + \sin(\tau)\mathbf{t}^1).
\eeaa
For the potential $\Theta$
\beaa
\Theta(\tau)=\Theta_0(\cosh(\tau)\mathbf{t}^1 + \sinh(\tau)\mathbf{t}^2;  
\cosh(\tau)\mathbf{t}^2 + \sinh(\tau)\mathbf{t}^1)
\eeaa
\paragraph*{Example 5}
Let $\Phi^1=\Psi^1$, $\Phi^2=\Psi^2$, $\Phi^0=1$.
We will consider this example only for the times entering  the SDSC system (\ref{SDCS3}),
all higher times of the hierarchy are considered to be zero.
Introduction of higher times leads to complicated extra terms. Lax-Sato equation 
of Orlov-Schulman symmetry  in terms of `plus' vector fields (\ref{OSc+}) read
\beaa
\partial_\tau\mathbf{\Psi}=
+\left(
\frac{1}{J_0}
\begin{vmatrix}
\Psi^1_\lambda & \Psi^2_\lambda\\
\Psi^1_y & \Psi^2_y
\end{vmatrix}
\right)_+
\partial_x \mathbf{\Psi}
-\left(\frac{1}{J_0}
\begin{vmatrix}
\Psi^1_\lambda & \Psi^2_\lambda\\
\Psi^1_x & \Psi^2_x
\end{vmatrix}
\right)_+
\partial_y \mathbf{\Psi}
\nn\\
-\left(
\frac{1}{J_0}
\begin{vmatrix}
\Psi^1_x & \Psi^2_x\\
\Psi^1_y & \Psi^2_y
\end{vmatrix}
\right)_+
\partial_\lambda \mathbf{\Psi}
+
\begin{pmatrix}
1
\\
0
\\
0
\end{pmatrix}
.
\eeaa
Directly calculating the coefficients, we obtain
\beaa
\partial_\tau 
\begin{pmatrix}
\Psi^0
\\
\Psi^1
\\
\Psi^2
\end{pmatrix}
=
(z\p_x +w p_y -\p_\lambda)
\begin{pmatrix}
\Psi^0
\\
\Psi^1
\\
\Psi^2
\end{pmatrix}
+
\begin{pmatrix}
1
\\
0
\\
0
\end{pmatrix},
\eeaa
and for the functions $f$, $F$, $G$ we get the symmetry in differential form
\beaa
\partial_\tau f=(z\p_x +w p_y) f,
\quad
\partial_\tau F=(z\p_x +w p_y) F,
\quad
\partial_\tau G=(z\p_x +w p_y) f,
\eeaa
which can be explicitely integrated and gives a Galilean transformation
\bea
f(\tau)&=&f_0(x+\tau z, z; y+\tau w, w),
\nn
\\
F(\tau)&=&F_0(x+\tau z, z; y+\tau w, w),
\nn
\\
G(\tau)&=&G_0(x+\tau z, z; y+\tau w, w).
\label{GalX}
\eea
This transformation is different from the Galilean transformation of previous examples,
as it combines lower and higher order independent variables, and  above we considered
combinations of the variables of the same order, but belonging to different
sets $\mathbf{t}^1$,  $\mathbf{t}^2$. Thus transformation (\ref{GalX}) 
resembles the Galilean transformation
for the Manakov-Santini hierarchy \cite{LVB-OS25}, where we have only one set of times 
of the hierarchy.
For the second heavenly equation (\ref{Heav}), we have the Galilean transformation for the
potential, $$\Theta(\tau)=\Theta_0(x+\tau z, z; y+\tau w, w).$$
\section{Orlov-Schulman symmetries in terms of the Riemann-Hilbert problem}
First we briefly describe the dressing scheme for the SDCS hierarchy,
following the work \cite{LVB25}.
A dressing scheme can be formulated
in terms of three-component nonlinear Riemann-Hilbert problem on the unit circle $S$
in the complex plane of the variable $\lambda$,
\bea
&&
\Psi^0_\text{out}=R_0(\Psi^0_\text{in},\Psi^1_\text{in},\Psi^2_\text{in}),
\nn\\
&&
\Psi^1_\text{out}=R_1(\Psi^0_\text{in},\Psi^1_\text{in},\Psi^2_\text{in}),
\nn\\
&&
\Psi^2_\text{out}=R_2(\Psi^0_\text{in},\Psi^1_\text{in},\Psi^2_\text{in}),
\label{RiemannSDCS}
\eea
where the functions 
$\Psi^0_\text{in}$,  $\Psi^1_\text{in}$, $\Psi^2_\text{in}$
are analytic inside the unit circle,
the functions 
$\Psi^0_\text{out}$,  $\Psi^1_\text{out}$, $\Psi^2_\text{out}$
are analytic outside the
unit circle and have an expansion of the form (\ref{form0}), (\ref{form1}),  (\ref{form2})
at infinity.
The functions $R_0$, $R_1$, $R_2$  are suggested to define a complex-analytic
diffeomorphism $\mathbf{{R}}\in\text{Diff(3)}$, and we call them
the dressing data. The problem
(\ref{RiemannSDCS}) implies the generating relation for
Lax-Sato equations (\ref{SD1}), (\ref{SD2}) (see \cite{BDM06,BDM07,LVB09})
\be
(J_0^{-1}\d \Psi^0\wedge \d \Psi^1\wedge \d \Psi^2)_-=0,
\label{analyticity0D}
\ee 
where the independent variables of the
differential include  the times $\mathbf{t^1}$,  $\mathbf{t^2}$ and $\lambda$.
Thus the Riemann problem (\ref{RiemannSDCS}) provides solutions for the SDSC
hierarchy (\ref{SD1}), (\ref{SD2}).

Let us consider in more detail, how the dynamics is introduced in terms of the dressing scheme.
Riemann-Hilbert problem (\ref{RiemannSDCS})
can be symbolically written in the form
\bea
\mathbf{\Psi}_\text{out}= \mathbf{R}(\mathbf{\Psi}_\text{in}),
\label{RiemannSD}
\eea
where $\mathbf{R}$ is a complex-analytic diffeomorphism, $\mathbf{R} \in \text{Diff(3)}$.
A natural way to introduce dynamics is to define the evolution of the dressing data
(diffeomorphism $\mathbf{R}$). Starting from one-parametric group of diffeomorphisms
$\mathbf{f}_\tau$, connected with some vector field $\hat {\mathcal{V}}$,
\bea
\hat {\mathcal{V}}=
{\mathcal{V}}^0(\Psi^0,\Psi^1,\Psi^2)\frac{\p~}{\p\Psi^0}
+{\mathcal{V}}^1(\Psi^0,\Psi^1,\Psi^2)\frac{\p~}{\p\Psi^1}
+{\mathcal{V}}^2(\Psi^0,\Psi^1,\Psi^2)\frac{\p~}{\p\Psi^2},
\label{vectorfieldSD}
\eea
we define
\bea
\mathbf{R}(\tau)=\mathbf{f_\tau}^{-1}\circ\mathbf{R}_0\circ\mathbf{f_\tau}
\label{evolution}
\eea
The times of the hierarchy $t_n$ correspond to commuting vector fields 
$\hat {\mathcal{V}}^1_{n}= (\Psi^0)^n \frac{\p~}{\p\Psi^1}$,
$\hat {\mathcal{V}}^2_{n}= (\Psi^0)^n \frac{\p~}{\p\Psi^2}$,
for which diffeomorphisms can be found explicitly,
\beaa
\Psi^0 \rightarrow \Psi^0, 
\quad 
\Psi^1 \rightarrow \Psi^1 + \sum_{n=0}^{\infty}t^1_n (\Psi^0)^n,
\quad
\Psi^2 \rightarrow \Psi^2 + \sum_{n=0}^{\infty}t^2_n (\Psi^0)^n,
\eeaa
and we arrive to Riemann-Hilbert problem of the form (\ref{RiemannSDCS}), where 
diffeomorphism $\mathbf{R}$ is independent of times, and dynamics is defined by the
`singular terms'  of $\Psi^1$, $\Psi^2$ (\ref{form1}),  (\ref{form2}),
in some similarity with the Baker-Akhiezer function (for which the singularity is multiplicative).
To this Riemann-Hilbert problem we can consistenly add evolution of the dressing data of the form
(\ref{evolution}), corresponding to some vector field (\ref{vectorfieldSD}).  The flows corresponding to
non-commuting vector fields do not commute. These flows represent Orlov-Schulmann symmetries
of the MS hierarchy in terms of the dressing scheme.

To obtain  Lax-Sato equations for the Orlov-Schulmann symmetries, 
one needs to add some extra
terms to the generating equation (\ref{analyticity0D}). They can be obtained by modifying the differentials
\beaa
&&
\d \Psi^0 \rightarrow \d \Psi^0+(\p_\tau \Psi^0 + (\hat {\mathcal{V}} \Psi^0))\d \tau,
\\
&&
\d \Psi^1 \rightarrow \d \Psi^1+(\p_\tau \Psi^1 + (\hat {\mathcal{V}} \Psi^1))\d \tau,
\\
&&
\d \Psi^2 \rightarrow \d \Psi^2+(\p_\tau \Psi^2 + (\hat {\mathcal{V}} \Psi^2))\d \tau.
\\
\eeaa
then generating equation (\ref{analyticity0D}) provides all the necessary extra relations.

In terms of infinitesimal symmetries,
\bea
&&
\delta_{\hat {\mathcal{V}}}\begin{pmatrix}
\Psi^0\\
\Psi^1\\
\Psi^2
\end{pmatrix}
=
\left(\left(
\frac{1}{J_0}
{\begin{vmatrix}
\hat {\mathcal{V}} \Psi^0& \hat {\mathcal{V}} \Psi^1& \hat {\mathcal{V}} \Psi^2\\
\Psi^0_\lambda & \Psi^1_\lambda & \Psi^2_\lambda \\
\Psi^0_y & \Psi^1_y & \Psi^2_y \\
\end{vmatrix}}
\right)_-\p_x
-
\left(\frac{1}{J_0}
{\begin{vmatrix}
\hat {\mathcal{V}} \Psi^0& \hat {\mathcal{V}} \Psi^1& \hat {\mathcal{V}} \Psi^2\\
\Psi^0_\lambda & \Psi^1_\lambda & \Psi^2_\lambda \\
\Psi^0_x & \Psi^1_x & \Psi^2_x \\
\end{vmatrix}}
\right)_-\p_y
\right.
\nn
\\&&
-
\left.
\left(\frac{1}{J_0}
{\begin{vmatrix}
\hat {\mathcal{V}} \Psi^0& \hat {\mathcal{V}} \Psi^1& \hat {\mathcal{V}} \Psi^2\\
\Psi^0_x & \Psi^1_x & \Psi^2_x \\
\Psi^0_y & \Psi^1_y & \Psi^2_y \\
\end{vmatrix}}
\right)_-\p_\lambda
\right)
\begin{pmatrix}
\Psi^0\\
\Psi^1\\
\Psi^2
\end{pmatrix}.
\label{OSVinfSD}
\eea 
For `plus' vector fields
\bea
&&
\delta_{\hat {\mathcal{V}}}\begin{pmatrix}
\Psi^0\\
\Psi^1\\
\Psi^2
\end{pmatrix}
=
\left(\left(-
\frac{1}{J_0}
{\begin{vmatrix}
\hat {\mathcal{V}} \Psi^0& \hat {\mathcal{V}} \Psi^1& \hat {\mathcal{V}} \Psi^2\\
\Psi^0_\lambda & \Psi^1_\lambda & \Psi^2_\lambda \\
\Psi^0_y & \Psi^1_y & \Psi^2_y \\
\end{vmatrix}}
\right)_+\p_x
+
\left(\frac{1}{J_0}
{\begin{vmatrix}
\hat {\mathcal{V}} \Psi^0& \hat {\mathcal{V}} \Psi^1& \hat {\mathcal{V}} \Psi^2\\
\Psi^0_\lambda & \Psi^1_\lambda & \Psi^2_\lambda \\
\Psi^0_x & \Psi^1_x & \Psi^2_x \\
\end{vmatrix}}
\right)_+\p_y
\right.
\nn
\\&&
+
\left.
\left(\frac{1}{J_0}
{\begin{vmatrix}
\hat {\mathcal{V}} \Psi^0& \hat {\mathcal{V}} \Psi^1& \hat {\mathcal{V}} \Psi^2\\
\Psi^0_x & \Psi^1_x & \Psi^2_x \\
\Psi^0_y & \Psi^1_y & \Psi^2_y \\
\end{vmatrix}}
\right)_+\p_\lambda
\right)
\begin{pmatrix}
\Psi^0\\
\Psi^1\\
\Psi^2
\end{pmatrix}
-\hat{\mathcal{V}}
\begin{pmatrix}
\Psi^0\\
\Psi^1\\
\Psi^2
\end{pmatrix}
\label{OSVinfSD+}
\eea 
Commutator of infinitesimal symmetries (\ref{OSVinfSD}) is a symmetry, 
corresponding to a commutator
of respective vector fields (\ref{vectorfieldSD}),
\beaa
[\delta_{\hat {\mathcal{V}}_1},\delta_{\hat {\mathcal{V}}_2}]=
\delta_{[\hat {\mathcal{V}}_2,\hat {\mathcal{V}}_1]},
\eeaa
and the linear map $\hat {\mathcal{V}}\mapsto \delta_{\hat {\mathcal{V}}}$ defines a 
homomorphism of a  Lie algebra of three-dimensional vector fields into the algebra
of symmetries of the MS hierarchy (compare \cite{Takasaki92,Takasaki95}).

Basic vector field 
$\hat{\mathcal{V}}={{\mathcal{V}}}^0(\Psi^0,\Psi^1,\Psi^2)\frac{\p~}{\p\Psi^0}$
gives a symmetry (\ref{OSVinfSD}) of the form (\ref{OSsd}) 
with 
$\Phi^0=-{\mathcal{V}}^0(\Psi^0,\Psi^1,\Psi^2)$, $\Phi^1=\Psi^1$,  $\Phi^2=\Psi^2$.
Respectively, a vector field 
$\hat{\mathcal{V}}={{\mathcal{V}}}^1(\Psi^0,\Psi^1,\Psi^2)\frac{\p~}{\p\Psi^1}$ 
corresponds to symmetry (\ref{OSsd})
with 
$\Phi^0={\mathcal{V}}^1(\Psi^0,\Psi^1,\Psi^2)$, $\Phi^1=\Psi^0$,  $\Phi^2=\Psi^2$,
and for a vector field
$\hat{\mathcal{V}}={{\mathcal{V}}}^2(\Psi^0,\Psi^1,\Psi^2)\frac{\p~}{\p\Psi^2}$ 
we have
$\Phi^0={\mathcal{V}}^2(\Psi^0,\Psi^1,\Psi^2)$, $\Phi^1=\Psi^1$,  $\Phi^2=\Psi^0$.
General vector field (\ref{vectorfieldSD}) leads to a linear combination of symmetries
of the form  (\ref{OSsd}). 

Representation of the Orlov-Schulman symmetries in the form (\ref{OSsd}) is convenient
to perform a volume-preserving reduction, corresponding to divergence-free vector fields.
For $J_0=1$, $\Phi^0=1$ vector fields in (\ref{OSsd}) are  divergence-free, and the symmetry
is compatible with the volume-preserving reduction $J_0=1$ of the hierarchy.

For the elementary examples of Orlov-Schulman symmetries that 
we considered above (Galilean transformations and scalings) 
vector fields $\hat {\mathcal{V}}$ have zero
or constant divergence, and they are compatible with the volume-preserving reduction.


\begin{thebibliography}{99}
\bibitem{LVB-OS25} 
L.V.Bogdanov, The Orlov–Schulman Symmetries of the Ma\-na\-kov-Santini Hierarchy, Lobachevskii J Math 46 (2025) 5753–-5762. 
https://doi.org/10.1134/S1995080225612913

\bibitem{OS}
A.Y. Orlov and  E.I.  Schulman, Additional symmetries for integrable equations and conformal algebra representation. Lett Math Phys 12 (1986) 171--179.

\bibitem{Orlov88}
A.Yu. Orlov, Vertex operators, $\dbar$-problems, symmetries, variational
indentities and Hamiltonian formalism for 2 + 1 integrable systems,
in: Plasma Theory and Nonlinear and Turbulent Processes in Physics
(World Scientific, Singapore, 1988).

\bibitem{GrinOrlov89}
P.G. Grinevich and  A.Yu. Orlov, Virasoro action on Riemann surfaces,
Grassmannians, det $\dbar_j$ and Segal Wilson $\tau$ function, in: Problems
of Modern Quantum Field Theory (Springer-Verlag, 1989).

\bibitem{Takasaki92}
T. Takasaki, T. Takebe,
SDiff(2) KP hierarchy,
Int. J. Mod. Phys. A (1992), 889--922.

\bibitem{MS06}
S. V. Manakov and P. M. Santini,
The Cauchy problem on the plane for the dispersionless
Kadomtsev-Petviashvili equation,
\textit{JETP Lett.} \textbf{83} (2006) 462--466.

\bibitem{MS07}
S.V. Manakov and P.M. Santini,
A hierarchy of integrable PDEs in 2+1 dimensions associated with
2-dimensional vector fields,
\textit{Theor. Math. Phys.} \textbf{152} (2007) 1004--1011.

\bibitem{MS08}
S.V. Manakov and P.M. Santini,
On the solutions of the dKP equation:
the nonlinear, Riemann-Hilbert problem, longtime behaviour,
implicit solutions and wave breaking,
\textit{J Phys. A: Math. Theor.} \textbf{41} (2008) 055204.

\bibitem{DFK15}M. Dunajski, E.V. Ferapontov and 
B. Kruglikov, On the Einstein-Weyl 
and conformal self-duality equations,
Journal of Mathematical Physics 56(8) (2015) 083501.



\bibitem{BDM06}
L. V. Bogdanov, V. S. Dryuma and S. V. Manakov,
On the dressing method for Dunajski anti-self-duality equation,
\texttt{nlin/0612046} (2006).

\bibitem{BDM07}
L. V. Bogdanov, V. S. Dryuma and S. V. Manakov,
Dunajski generalization of the second heavenly equation:
dressing method and the hierarchy,
\textit{J Phys. A: Math. Theor.} 
\textbf{40} (2007) 14383--14393.

\bibitem {FK25} E.V. Ferapontov and B. Kruglikov, Involutive Scroll Structures on Solutions of 4D Dispersionless Integrable Hierarchies,  Commun. Math. Phys. 406 (2025) 299.
https://doi.org/10.1007/s00220-025-05479-z

\bibitem{Dun02}  M. Dunajski, 
Anti-self-dual four–manifolds with a parallel real spinor,
{Proc. Roy. Soc. Lond. A}
{458} 1205 (2002) 1205

\bibitem{Pleb}J.F. Pleba\'nski, 
Some solutions of complex Einstein equations,
J. Math. Phys. {16} (1975)  2395--2402 

\bibitem{Takasaki89}
K. Takasaki, 
An infinite number of hidden variables in hyperKähler metrics,
Journal of Mathematical Physics 30 (1989) 1515--1521. 

\bibitem{Takasaki90}
K. Takasaki, 
Symmetries of hyper-Kähler (or Poisson gauge field) hierarchy,
Journal of Mathematical Physics 31 (1990) 1877--1888.

\bibitem{LVB09} L. V. Bogdanov, A class of multidimensional integrable hierarchies 
and their reductions,
\textit{Theoretical and Mathematical Physics}, 
\textbf{160}(1)  (2009)  888--894.

\bibitem{Takasaki95}
K. Takasaki,
Symmetries and tau function of higher dimensional dispersionless integrable hierarchies,
J. Math. Phys. 36 (1995) 3574--3607.

\bibitem{LVB25} L.V. Bogdanov, Differential and other reductions of the self-dual conformal structure equations, Physica D: Nonlinear Phenomena
483 (2025) 135001.

\end{thebibliography}
\end{document}